\documentclass[12pt]{article}
\usepackage{setspace}
\usepackage[utf8]{inputenc} % usually not needed (loaded by default)
\usepackage[T1]{fontenc}
\usepackage{amsthm}
\usepackage{amsmath}               
{
	\theoremstyle{plain}

	\theoremstyle{remark}
	
	\theoremstyle{definition}
	
	\newtheorem{algorithm}{Algorithm}
}              
\usepackage{mathtools}
\usepackage{amssymb}
\usepackage{url}
\usepackage{graphicx}
\usepackage{dsfont}
\usepackage{natbib}
\newcommand{\blind}{0}
\usepackage[english]{babel}

\usepackage{dutchcal}
\usepackage{float}

\usepackage{color}
\usepackage{epsfig}
\usepackage{parskip}
\usepackage{multirow}
\usepackage{pdflscape}
\usepackage{verbatim}
\usepackage{caption}
\usepackage{hyperref}

\newcommand\myeq{\mathrel{\overset{\makebox[0pt]{\mbox{\normalfont\tiny\sffamily def}}}{=}}}

\author{Anwesha Bhattacharyya and Yves Atchade}

\begin{document}
	
	\def\spacingset#1{\renewcommand{\baselinestretch}%
		{#1}\small\normalsize} \spacingset{1}

	%%%%%%%%%%%%%%%%%%%%%%%%%%%%%%%%%%%%%%%%%%%%%%%%%%%%%%%%%%%%%%%%%%%%%%%%%%%%%%
	
	\if0\blind
	{
		\title{\bf A Bayesian Analysis of Large Discrete Graphical Models}
		\author{Anwesha Bhattacharyya\thanks{
				 anwebha@umich.edu}\hspace{.2cm}\\
			Department of Statistics, University of Michigan\\
			and \\
			Yves Atchade\thanks{atchade@bu.edu} \\
			Department of Statistics, Boston University}
		\maketitle
	} \fi
	
	\if1\blind
	{
		\bigskip
		\bigskip
		\bigskip
		\begin{center}
			{\LARGE\bf A Bayesian Method for large Discrete Graphical models}
		\end{center}
		\medskip
	} \fi

	\vfill
	\footnotetext{\textit{This work is partially supported by the NSF grant DMS 1513040.
			}}
	\newpage
	%\spacingset{1.2} % DON'T change the spacing!
\doublespacing
	
		\bigskip

		\begin{abstract}
			This work introduces a Bayesian methodology for fitting large discrete graphical models with spike-and-slab priors to encode sparsity. We consider a quasi-likelihood approach that enables node-wise parallel computation resulting in reduced computational complexity. We introduce scalable MCMC algorithms for sampling from the quasi-posterior distribution which enables variable selection and estimation simultaneously. We present extensive simulation results to demonstrate scalability and accuracy of the method. We also analyze the 16 Personality Factors (PF) dataset to illustrate performance of the method. A Matlab implementation of the algorithms is also provided as support material. 
		\end{abstract}
	\noindent%
	\newpage
 
	\section{Introduction}
	\label{sec:intro}
	 In this paper we are primarily interested in model-based inference of large undirected networks from observed data at each node, in settings where the data can take only finitely many values. This is motivated by the widespread availability of this type of data in areas of psychology,  image processing, computer science, social sciences, bio-informatics, to name a few. For instance, \cite{banetal:2008} used an Ising model to find association between US senators from their binary voting records. \cite{ekeberg} used a Potts model to predict contact between amino acids in protein chains. In \cite{epskamp:2017,epskampetal}, the authors worked the reader through   statistical procedures for estimating psychological networks in personality research. 
	The Ising model \cite{ising} was originally formulated in the Physics literature as a simple model for interacting magnetic spins on a lattice. The Potts model is a generalization of the Ising model in which spins can take more than two values with more complex dependencies. These models are widely used in the applications for  teasing out direct and undirected dependencies between large collections of variables. The purpose of this work is to construct robust and scalable bayesian  procedures for fitting these models.
	
	Bayesian statistics has two built-in features that are in growing demand in the applications: a) the ability to incorporate existing knowledge in new data analysis  (\cite{r9,r10,r11}), and b) a simple mechanism for uncertainty quantification in the inference. However these features come at a hefty computational price: most existing Bayesian methods for fitting graphical models do not scale well as the number of nodes in the graph grows, despite   the recent progress with Gaussian graphical models (\cite{dobra:2011,khondker:2013,peterson:2015,ban:ghoshal}).  The computational challenge only intensifies when dealing with discrete graphical models (Ising and Potts models). Here  a full Bayesian treatment leads to the so-called doubly-intractable posterior distributions  for which specialized MCMC algorithms are needed (\cite{r5,murray:etal:06,r7}). However these algorithms do not scale well  when dealing with large graphical models.

	In the frequentist literature there is a long history of fitting discrete graphical models using (quasi/pseudo)-likelihood methods instead of the full likelihood (dating back at least to \cite{r2}; see also \cite{guyon:95}). In fact, quasi-likelihood methods have become the de facto approach in the frequentist literature when dealing with large graphical models (\cite{buhlmann:meinshausen:06,hoefling:etal:09,ravikumaretal10,guo:etal:2015,roy:etal:17}). 
	 
	In the Bayesian framework, the idea of using a non-likelihood function to carry out  inference has been adopted at a more slower pace, but is currently in growing popularity   (\cite{chernozhukov:hong03,jiang:tanner08,liao:jiang:11,yang:he:2012,kato:2013,li:jiang:14,atchade:2017,atchade:2018}). For instance  in \cite{atchade:2018} a  quasi-likelihood approach is developed to fit Gaussian graphical models in the Bayesian framework at a scale unmatched by fully Bayesian alternatives. The crux of the method is the use of a  product-form pseudo-likelihood function (as used in the frequentist literature) that makes it possible to split the resulting quasi-posterior distribution  into a product of linear regression Bayesian posterior distributions. Significant reducing in computational costs is then achieved by deploying this  approach on a multi-core computer system.  In this paper we extend the same strategy to the analysis of discrete graphical models.

	We focus on settings where the underlying network is sparse. Under this assumption, the inference problem is commonly approached by introducing an auxiliary selection variable that represents the structure of the network. We then follow a well-established practice in the Bayesian literature that  imposes a spike-and-slab prior distribution jointly on the network parameter and the network structure (\cite{mitchell:beuchamp,george:mcculloch,narisetty:he,atchade:2019}).  More precisely, we actually follow here the sparse model spike and slab prior of \cite{atchade:2019}, a computationally efficient version of the standard spike and slab prior of \cite{george:mcculloch}.

	In summary, in this paper  we put together a scalable Bayesian methodology for fitting high-dimensional Potts or Ising models. Our method simultaneously estimates the model parameters and the underlying structure of the graph.  We develop  scalable Markov Chain Monte Carlo (MCMC) algorithms that can be implemented in parallel thus significantly reducing the computational cost of the method. Our method not only provides us with robust point estimates but also gives us credible intervals for the model parameters. We extensive tested the method using simulated data, and illustrates its practical applicability using an example from personality research.
		    
	 The rest of the article is organized as follows: Section \ref{model} introduces the methodology. In section \ref{mcmc}, we propose two scalable  MCMC algorithms to deal with the resulting quasi-posterior distribution. Section \ref{sim} illustrates the performance of our method through simulation results. Finally, in section \ref{sec:data}, we present an application of our method in the context of psychological data through the analysis of the 16 Personality Factors (16PF) dataset.
\vspace{5mm}
\section{Quasi-posterior distribution of the Potts model under spike and slab prior}
\label{model}
 An $m$-colored Potts model parametrized by a sparse symmetric matrix $\theta$ is a probability mass function on $\mathcal{Z} = \{0,1,\cdots, m-1\}^p$ given by
\begin{equation}
\label{PoMo}
f(z_1, \cdots z_p|\theta) = \frac{1}{\Psi(\theta)}\exp{\Big \{\sum_{r=1}^{p}\theta_{rr}C(z_r) + \sum_{r=1}^{p}\sum_{j< r }^{p}\theta_{rj}C(z_r,z_j)\Big\}}.
\end{equation} 
Here $\Psi(\theta) = \sum_{z \in \mathcal{Z}}\exp{\Big \{\sum_{r=1}^{p}\theta_{rr}C(z_r) + \sum_{r=1}^{p}\sum_{j< r }^{p}\theta_{rj}C(z_r,z_j)\Big\}}$ is the normalizing constant. The mean field function $C(.)$ describes the marginal information on $z_r$ while the coupling function $C(.,.)$ as suggested by the name describes the interaction between $z_r$ and $z_j$. A special case of \eqref{PoMo} is the Ising Model where $m$ is 2, and hence $\mathcal{Z} = \{0,1\}^p$. In case of the Ising model the mean field and the coupling functions are typically taken as identity ($C(z_r) = z_r$) and multiplicative ($C(z_r,z_j) = z_rz_j$) respectively.\label{l1}\\

The problem of interest in this work is the estimation and recovery of the sparse matrix $\theta$ based on $n$ sample observations $\{z^i\}_{i=1}^n$, where $z^i = (z^i_1, \cdots, z^i_p) \in \mathcal{Z}$ is the $i_{th}$ observation. We use $Z\in\{0,\ldots,m-1\}^{n \times p}$ to denote the matrix of observations, where the $i$-th row of $Z$ is $z^i$. The likelihood of $\theta$ can then be expressed as
\[\mathcal{L}^n(\theta|Z) = \prod_{i=1}^{n} f(z^i|\theta) = \prod_{i=1}^{n}\frac{1}{\Psi(\theta)}\exp{\Big \{\sum_{r=1}^{p}\theta_{rr}C(z_r^i) + \sum_{r=1}^{p}\sum_{j< r }^{p}\theta_{rj}C(z_r^i,z_j^i)\Big\}}.\]  
In a high-dimensional setting (typically $p>n, \frac{\log(p)}{n} \to 0$), likelihood based inference on $\theta$ is computationally intractable because of the normalization constant $\Psi(\theta)$. Note that,
the number of summands in the normalizing constant $\Psi(\theta)$ is exponential in $p$, and quickly blows up for even moderate values of $p$. 

\subsection{Quasi(Pseudo)-likelihood}
Following an approach widely adopted in the high-dimensional frequentist literature, we explore the use of quasi(pseudo)-likelihoods in the Bayesian treatment of discrete graphical models. The conditional distribution for the $r_{th}$ node (given all other nodes) in a Potts model for the $i_{th}$ observation $z^i$ can be written as 
\begin{equation} \label{eq1}
f(z^i_r | z^i_{\setminus r},\theta_r) = \frac{1}{\Psi_r^i(\theta_r)}\exp{\Big \{\theta_{rr}C(z^i_r) + \sum_{j\neq r}\theta_{rj}C(z^i_r,z^i_j)\Big\}},
\end{equation} 
where $z^i_{\setminus r} = (z^i_1, \cdots, z^i_{r-1},z^i_{r+1},\cdots z^i_p)'$ and $\theta_r = (\theta_{r1}, \cdots, \theta_{rp})'$ is the $r_{th}$ column of $\theta$. The normalizing constant of this conditional distribution is given by
\[\Psi_r^i(\theta_r) = \sum_{s=0}^{m-1}\exp\Big(\theta_{rr}C(s) + \sum_{j \neq r}\theta_{rj}C(s,z_j^i)\Big).\] 
Computing $\Psi_r^i(\theta_r)$ requires $O(p\times m)$ units of operations and hence is scalable when $m$ is small. We denote the $r_{th}$ conditional log-likelihood as 
\[\ell_{r}^{n}(\theta_r|Z) = \sum_{i=1}^{n}\bigg[\theta_{rr}C(z^i_r) + \sum_{j \neq r}\theta_{rj}C(z^i_r, z^i_j)-\log \bigg(\Psi_r^i(\theta_r)\bigg)\bigg].\] 
Following \cite{buhlmann:meinshausen:06,ravikumaretal10},  we consider the log pseudo-likelihood of $\theta$ given by
\begin{equation}\label{psl}
\ell^{n}(\theta \vert Z) = \sum_{r=1}^{p}\ell_{r}^{n}(\theta_r \vert Z).
\end{equation}
Note that the ability to write the log pseudo-likelihood $\ell^n(\theta \vert Z)$ as a sum of log conditional likelihoods $\ell_{r}^{n}(\theta_r|Z)$ allows us to transform the inference on $\theta\in\mathbb{R}^{p\times p}$ into $p$ separable sub-problems on $\mathbb{R}^p$. Parallel treatment of each of these regression problems when deploying a multi-core computer increases computational efficiency but comes at a cost of loss in symmetry in the estimated matrix $\theta$. We get two estimates for each component $\theta_{ij}$ from the computations involving nodes $i$ and $j$ respectively. Following \cite{buhlmann:meinshausen:06} we resolve this issue at the post-inference stage by taking an aggregate of the two estimates which shall be discussed in details in the later sections.  

\subsection{Spike and slab prior}

To take advantage of the factorized form of the pseudo-likelihood function from (\ref{psl}) we will assume in our prior distribution that the columns of $\theta$ are independent. We note that it is a common practice in Bayesian data analysis to ignore unknown dependence structure among parameters in the prior distribution when dealing with multivariate parameters. These dependences are then learned from the data in the posterior distribution. As mentioned before, the lack of symmetry is dealt with at the post-inference stage.

 As a prior distribution for $\theta_r$ we propose to use a relaxed form of the spike and slab prior (\cite{mitchell:beuchamp},\cite{george:mcculloch}). More specifically, for each parameter $\theta_r\in\mathbb{R}^p$, $r = 1,\cdots ,p$, we introduce a selection parameter $\delta_r = (\delta_{r1}, \cdots , \delta_{rp}) \in \Delta$, where $\Delta = \{0,1\}^p$. We assume that the component of $\delta_r$ have independent Bernoulli prior distributions, so that the joint distribution of $\delta_r$ writes
\begin{equation}
\omega_{\delta_r} = \prod_{j=1}^{p}q^{\delta_{rj}}(1-q)^{\delta_{rj}} \ ; \  q = p^{-(u+1)} \ ; \ u>0  \label{selprior}
\end{equation}
where $u$ is a hyper-parameter. The conditional distribution of $\theta_r$ given $\delta_r$ is given by
 \begin{align}
\nonumber \theta_{rj}&\vert \{\delta_{rj} = 1\} \sim \textbf{N}(0,\rho); \ \rho>0\\
\theta_{rj}&\vert \{\delta_{rj} = 0\} \sim \textbf{N}(0,\gamma); \ \gamma>0,
  \end{align}

We introduce the notations $\theta_{r\delta_r} = (\theta_{rj} \ s.t. \  \delta_{rj}=1) \in \mathbb{R}^{\|\delta_r\|_1}$, $\delta_r^c = 1 - \delta_r$, $\|z\|_1 = \sum_{j=1}^{p}|z_j|$ and $\|z\|_2 = \sqrt{\sum_{j=1}^{p}z_j^2}$. Using this notation, and writing $\delta=(\delta_1,\ldots,\delta_p)$, $\theta=(\theta_1,\ldots,\theta_p)$, % and $\rho = (\rho_1,\rho_2,\cdots, \rho_p) \in \mathbb{R}^p$,
 the joint prior distribution of $(\delta,\theta)\in \Delta^p\times\mathbb{R}^{p\times p}$ is given by
\[\pi(\delta, d\theta) = \prod_{r=1}^p \pi(\delta_r, d\theta_r) .\]
 The prior distribution $\pi(\delta_r, d\theta_r)$ on $\Delta \times \mathbb{R}^p$ can be written as
\begin{equation} 
\pi(\delta_r, d\theta_r) \propto \omega_{\delta_r}\left(2\pi\rho\right)^{-\frac{||\delta_r||_1}{2}}(2\pi\gamma)^{\frac{||\delta_r||_1}{2}}\exp\left(-\frac{1}{2\rho}\sum_{j:\;\delta_{rj}=1}\theta_{rj}^2- \frac{1}{2 \gamma}\sum_{j:\;\delta_{rj}=0} \theta_{rj}^2\right)d\theta_r. \label{ssprior}
\end{equation}
 
\subsection{Quasi-posterior distribution}
Following \cite{atchade:2019}, we combine the  prior distribution in (\ref{ssprior}) together with the pseudo-likelihood $\ell_r(\cdot\vert Z)$ and consider the quasi-posterior distribution for the $r$-th column of $\theta$ on $\Delta \times\mathbb{R}^p$ given by
\begin{multline}
\label{postcon}
\Pi_{n}(\delta_r, d\theta_r\vert Z) 
\propto \omega_{\delta_r}\left(\frac{\sqrt{\gamma}}{\sqrt{\rho}}\right)^{||\delta_r||_1}\exp\left(\ell_r^n(\theta_{r\delta_r}\vert Z)-\frac{1}{2 \rho}\sum_{j:\;\delta_{rj}=1}\theta_{rj}^2- \frac{1}{2 \gamma}\sum_{j:\;\delta_{rj}=0} \theta_{rj}^2\right)d\theta_r.
\end{multline} 
Note the use of $\theta_{r\delta_r}$ (the sparsified version of $\theta_r$)  in the quasi-likelihood. Although we use the same standard Gaussian-Gaussian spike-and-slab prior (as in for instance \cite{george:mcculloch}, \cite{narisetty:he}), the quasi-posterior in (\ref{postcon}) differs from those considered in the aforementioned paper due to the sparsification of $\theta_{r\delta_r}$ in the quasi-likelihood.  The idea was introduced in \cite{atchade:2019} to facilitate computation and more closely  approximate the  quasi-posterior distribution  obtained from spike-and-slab with point-mass at the origin. The contraction  properties of (\ref{postcon})  are analyzed in \cite{atchade:2019}. We multiplicatively combine these $p$ quasi-posterior distributions to obtain the full quasi-posterior distribution on $(\delta,\theta)$ given by
\begin{equation}
\label{post}
\Pi_n(\delta,d\theta|Z) = \prod_{r=1}^{p}\Pi_{n}(\delta_r, d\theta_r|Z).
\end{equation}

\subsection{Choice of hyper-parameters }
\label{hyperparam}
The behavior of (\ref{postcon}) depends by and large on the choice of the hyper-parameter $\gamma,\rho$ and $u$. We refer the readers to \cite{atchade:2019} for a detailed discussion. In our algorithms we set $q$ in \eqref{selprior}  at $q = \frac{1}{p^{1+u}}$, for some constant $u>0$.  We have found that the inference is typically very robust to any choice of $u$ between $1$ and $2$.

 The hyper-parameter $\gamma$ is the prior variance of the inactive component, whereas $\rho$ is the prior variance of the active components.  We follow \cite{atchade:2019}, and for positive constants $c_0,c_1$, choose 
 \[\gamma = \frac{c_0}{\max(n,p)},\;\;\mbox{ and }\;\;\rho = c_1 \sqrt{\frac{n}{\log(p)}}.\]

\subsection{Post estimation symmetrization}
As mentioned above our procedure can lead to  two different set of estimates $\hat{\theta}_{ij}$ and $\hat{\theta}_{ji}$ for the same parameter $\theta_{ij}$. For the sake of interpretation it is useful to provide a single estimate and credible interval. We propose a post-estimation symmetrization resulting in a singular estimate 
\begin{equation}
\tilde{\theta}_{ij} =  \frac{\hat{\theta}_{ij} + \hat{\theta}_{ji}}{2}.
\label{strength}
\end{equation} 

Similarly, the credible region corresponding to the parameter $\theta_{ij}$ is constructed as union of the 95\% credible intervals $\theta_{ij}$ and $\theta_{ji}$. Taking the union is a conservative approach as opposed to taking the intersection. However it always provides a concrete interval or set unlike the intersection in which case the credible intervals may be too short or in some cases even result in null set. A more direct inference on the presence of edge between nodes $i$ and $j$ can be made from the indicator variable $\delta_{ij}$. In the same spirit as above we estimate $\delta_{ij}$ using

\begin{equation}
\tilde p_{ij} = P(edge \ between \ node \ i \ and \ j\vert Z) = \frac{1}{2}\left(\hat{P}(\delta_{ij} = 1\vert Z)+\hat{P}(\delta_{ji} = 1\vert Z)\right).
\label{edge}
\end{equation}

\section{MCMC Sampling Algorithms}
\label{mcmc}

In this section we shall discuss in details the construction of Markov Chain Monte Carlo (MCMC) algorithms to draw Monte Carlo samples from the posterior distribution \eqref{post}. By virtue of independence, it is enough to draw sample for each of the joint variable $(\theta_r,\delta_r)$. Large efficiency gain is possible by performing these simulations in parallel. In general
we adopt a Metropolis-Hastings within Gibbs approach to create our samplers.

We describe in Section \ref{mala} a general Metropolis Adjusted Langevin Algorithm (MALA) to sample from \eqref{post}. In case of Ising model, one can also take advantage of the fact that the conditional distributions are  logistic regression models and employ the Polya-Gamma(PG) sampler  of \cite{pg:polson} for sampling (Section \ref{polya-gamma}).  We compare the two schemes in Section \ref{sec:comp:algo}. 

\subsection{A Metropolis Adjusted Langevin sampler}\label{mala}
The algorithm updates the active components $\theta_{r\delta_r}$ given $(\delta_r,\theta_{r\delta_r^c})$, then updates the inactive components $\theta_{r\delta_r^c}$ given $(\delta_r,\theta_{r\delta_r})$, and finally updates $\delta_r$ given $(\theta_r)$. Here we have used the notations $\theta_r = [\theta_{r\delta_r},\theta_{r\delta_r^c}]$, where $\theta_{r\delta_r}$ regroups the components of $\theta_r$ for which $\delta_{rj} = 1$, and $\theta_{r\delta_r^c}$ regroups the remaining components. We refer the reader to \cite{robert:casella,liu} for an introduction to basic MCMC algorithms.\\

\noindent\textbf{Update of active parameters}\\
Suppose that $\delta_r$ is such that $0 < ||\delta_r||_1 < p$.
We update $\theta_{r\delta_r}$ by a Metropolis Adjusted Langevin Algorithm (\cite{atchade:2006}). Other algorithms including Hamiltonian Monte Carlo could be used as well. We define
\begin{equation}
h(\delta_r,\theta_r;z) = \Big[\ell_r^n( \theta_{r\delta_r}|z) - \frac{1}{2\rho}||\theta_{r\delta_r}||_2^2  - \frac{1}{2 \gamma}||\theta_{r\delta^c_r}||_2^2\Big]. \label{expon}
\end{equation}
The function $\theta_r \to h(\delta_r,\theta_r;z)$ has a gradient given by
\[\nabla_{\theta_r}h_\gamma(\delta_r,\theta_r;z) = \nabla_{\theta_{r\delta_r}}\ell_{r}^n(\theta_{r\delta_r}|z) - \frac{1}{\rho}\theta_{r\delta_r} - \frac{1}{\gamma}\theta_{r\delta^c_r} .\]
 Following (\cite{atchade:2006}), we further truncate the  gradient by introducing
 \begin{equation}
 G(\delta_r,\theta_r; z) \myeq \frac{c}{c \vee \|\nabla_{\theta_r}h(\delta_r,\theta_r;z)\|_2}\nabla_{\theta_r}h(\delta_r,\theta_r;z),\label{langrad}
 \end{equation}
 for some positive constant $c$, where $a\vee b = \max(a,b)$.
 We update (one at the time) the selected components of $\theta_r$ as follows. Given $j$ such that $\delta_{rj} = 1$, we propose
\begin{equation}\label{eq:langevin:update}
\theta_{rj}^{prop}|\theta_r \sim \mathbf{N}\Big(\theta_{rj} + \frac{\sigma}{2}[ G(\delta_r,\theta_r;z)]_j, \sigma^2 \Big),\end{equation}
where $\sigma$ is some constant step size and $[G(\delta_r,\theta_r,\rho_r;z)]_j$ represents the $j_{th}$ component of $G(\delta_r,\theta_r;z)$. Let $g(\theta_{rj}^{prop}|\theta_r)$ denote the density of the proposal distribution in \eqref{eq:langevin:update}. We also define $\theta_r^{prop} = [\theta_{r1},\cdots\theta_{r(j-1)}, \theta_{rj}^{prop},\theta_{r(j+1)},\cdots \theta_{rp}]$ and the acceptance probability as
\begin{equation}
\mathsf{Acc}_{rj} = \min\left(1,\frac{g(\theta_{rj}|\theta_r^{prop})}{g(\theta_{rj}^{prop}|\theta_r)}\times \frac{\Pi_n(\delta_r,\theta_r^{prop}|Z)}{\Pi_n(\delta_r,\theta_r|Z)}\right).
\label{accactive}
\end{equation}
With probability $\mathsf{Acc}_{rj}$ we set $\theta_{rj} = \theta_{rj}^{prop}$, and with probability $1 - \mathsf{Acc}_{rj}$, we do nothing.   In our simulations the step size $\sigma$ is kept constant. Alternatively, it can also be updated for each $\theta_{rj}$ in the spirit of an adaptive MCMC scheme if so desired.

Finally we note that, under sparse prior the number of active parameters in each node is small. Hence the active parameters at a node can be updated one by one without loss in computational efficiency.

\noindent\textbf{Independent  update for inactive parameters}\\
Note that for the stated posterior distribution \ref{postcon}, given $\delta_r$, the inactive components $\theta_{r\delta_r^c}$ can be updated from their full conditional distribution given by
\begin{equation}
\theta_{r\delta_r^c} \sim \mathbf{N}(\mathbf{0},\gamma I_{p - \|\delta_r\|_1}).\label{indep}
\end{equation}
\noindent\textbf{Bernoulli sampler for selection parameters}\\
Equation \eqref{postcon} is used to derive the one by one Gibbs update of the $\delta_{rj}$'s. For each $j = 1, \cdots p$, we define $\check{\delta}_r = (\delta_{r1},\cdots, \delta_{r (j-1)},\delta_{rj}^c,\delta_{r(j+1)},\cdots,\delta_{rp})$ and set
\begin{equation}
 \tau_{rj} = \min\left(1, \frac{(\frac{q}{1-q})^{\|\check{\delta}_r\|_0}(\frac{\gamma}{\rho})^{\frac{\|\check{\delta}_r\|_0}{2}}e^{h(\check{\delta}_r,\theta_r;z)}}{(\frac{q}{1-q})^{\|\delta_r\|_0}(\frac{\gamma}{\rho})^{\frac{\|\delta_r\|_0}{2}}e^{h(\delta_r,\theta_r;z)}}\right) \label{delmala}
\end{equation}
We change $\delta_{rj}$ to $\delta_{rj}^c$ based on a flip of probability $\tau_{rj}$

\medskip

The overall MCMC algorithm, hereafter referred to as MALA can be summarized as follows.

\begin{algorithm}\label{algo:mala} \textbf{MALA} sampler\\
For each node $r\in \{1,\cdots,p\}$ do the following.
\begin{enumerate}
	\item Initialize with $(\theta_r^{(0)},\delta_r^{(0)})$
	\item At the $t$-th iteration, given $\delta_r^{(t-1)} =\check{\delta}$ and $\theta_r^{(t-1)} = \check\theta$, do
	\begin{enumerate}
	\item For each $j$ such that $\check\delta_j=1$, we update $\check\theta_j$ using the MALA algorithm described in (\ref{eq:langevin:update}) and (\ref{accactive}). 
	\item Update $\check{\theta}_{\check{\delta}^c} \sim \mathbf{N}(0,\gamma I_{p - \|\check{{\delta}}\|_0})$
	\item Set $\theta_r^{(t)} = \check{\theta}$. For each $j$ in $\{1,\ldots,p\}$, we update
				 $\check{\delta}_j$  based on a binary flip of probability $\tau_{rj}$ as defined in \eqref{delmala}. Set $\delta_r^{(t)} = \check{\delta}$.
	
\end{enumerate}
\end{enumerate}
\end{algorithm}
\subsection{A Polya-Gamma sampler for Ising models}\label{polya-gamma}

The Polya-Gamma sampler is a data-augmentation technique which introduces latent Polya-gamma variables to obtain an efficient Gibbs sampler for Bayesian logistic regression (\cite{pg:polson}). To see how this is used here, note that the conditional posterior of the active parameters for the $r_{th}$ node is 
\begin{equation}
\Pi_n(\theta_{r\delta_r}|\delta_r,\theta_{r\delta_r^c};Z) \propto \exp\left(\ell_r^n(\theta_{r\delta_r}|Z) - \frac{1}{2\rho}\sum_{j:\delta_{rj = 1}}\theta_{rj}^2\right),\label{posac}
\end{equation}
which is the same as the posterior distribution in a logistic regression of  variable $z_r$ over the variables $z_j$ for which $\delta_{rj}=1$, $j\neq r$, using all available data samples. Given $r,\delta_r$, we write $x(r)^{(i)}_{\delta_r} = (z_1^{(i)},\cdots,z_{r-1}^{(i)},1,z_{r+1}^{(i)},\cdots,z_p^{(i)})_{\delta_r} \in \{0,1\}^{\|\delta_r\|_1}$ , $Z_r = (z_r^{(1)},\cdots,z_r^{(n)})' \in \{0,1\}^n$ and use $X(r)_{\delta_r} \in \{0,1\}^{n \times \|\delta_r\|_1} $ to denote the matrix of $n$ observations $\{x(r)^{(i)}_{\delta_r}\}_{i = 1}^n$.  

Hence to sample from \eqref{posac} we follow a Gibbs update of first drawing independently Polya-Gamma random variables using
\begin{equation}
W_i|\theta_{r\delta_r} \sim \mathbf{PG}(1 ,|\langle x(r)^{(i)}_{\delta_r},\theta_{r\delta_r}\rangle|) ; \  i = 1,\cdots ,n \label{pgup1}
\end{equation} 
 Note that $\langle a,b\rangle$ denotes the inner product between two vectors $a,b$. The second step is to update $\theta_{r\delta_r}$ given these Polya-Gamma variables using
 \begin{eqnarray}
 \theta_{r\delta_r} &\sim & \mathbf{N}(\mu,\Sigma)\label{pgup2}\\
 \mu &= & \Sigma\left(X(r)^T_{\delta_r}(Z_r - \frac{1}{2}1_n)\right)\\
 \Sigma &= & \left(X(r)^T_{\delta_r}\Omega X(r)_{\delta_r} + \frac{1}{\rho}I_{\|\delta_r\|_0}\right)^{-1}\\
 \Omega &=& \textsf{diag}(W_1,\cdots, W_n)\label{pgup3}
 \end{eqnarray}
 
\medspace
 	
\noindent\textbf{Independent  update for inactive parameters}\\
As in \eqref{indep} given $\delta_r$, the inactive components $\theta_{r\delta_r^c}$ can be updated independently and simultaneously from $\mathbf{N}(\mathbf{0},\gamma I_{p - \|\delta_r\|_1})$

\noindent\textbf{Bernoulli sampler for selection parameters}\\
As before, \eqref{postcon} is used to derive the one by one Gibbs update of the $\delta_{rj}$'s. For the Polya-Gamma (PG) sampler, the calculations of the Bernoulli probability of the update can be simplified. For each $j = 1, \cdots p$, we define 
\begin{multline}
\tau_{rj} = \log\left(\frac{1-q}{q}\right) - \frac{1}{2}\log\left(\frac{\gamma}{\rho}\right) + \frac{1}{2}\left(\frac{1}{\rho} - \frac{1}{\gamma}\right)\theta_{rj}^2 -\frac{1}{2}\Big([X(r)]'_{\cdot j}\Omega [X(r)]_{\cdot j}\Big)\theta_{rj}^2\\
-\theta_{rj}\Bigg\langle[X(r)]_{\cdot j},\left(Z_r - \frac{1}{2}1_n\right)\Bigg\rangle - \Big\langle\theta_{r\delta_r},[X(r)]'_{\cdot j}\Omega X(r)_{\delta_r}\Big\rangle
\label{taupg}
\end{multline}
where $X(r)$ denotes the full matrix $X(r)_{\mathbf{1}_p}$ and $[X(r)]_{\cdot j}$ denotes the $j_{th}$ column of $X(r)$.

When $\delta_{rj} = 1$ we flip it to $0$ with probability $\min(1,e^{\tau_{rj} + \theta_{rj}[X(r)]'_{\cdot j}\Omega [X(r)]_{\cdot j}})$. On the other hand if $\delta_{rj} = 0$ we flip it to $1$ with probability $\min(1,e^{-\tau_{rj}})$.

\medskip

The Polya-Gamma MCMC algorithm (hereafter PG sampler) can be summarized as follows.

\begin{algorithm}\label{algo:polya-gamma} \textbf{PG} sampler \\
	For each node $r\in \{1,\cdots,p\}$ do the following.
	\begin{enumerate}
		\item Initialize with $(\theta_r^{(0)},\delta_r^{(0)})$
		\item At the $t$-th iteration, given $\delta_r^{(t-1)} =\check{\delta}$ and $\theta_r^{(t-1)} = \check\theta$, do
		\begin{enumerate}
			\item we update $\check\theta_{\check{\delta}}$ using the Polya-Gamma algorithm described in (\ref{pgup1} - \ref{pgup3}). 
			\item Update $\check{\theta}_{\check{\delta}^c} \sim N(0,\gamma I_{p - \|\check{{\delta}}\|_0})$

			\item Set $\theta_r^{(t)} = \check{\theta}$. For each $j$ in $\{1,\ldots,p\}$\\
			 \textbf{IF}  $\check{\delta}_j = 1$\\
			  \indent we flip it to $0$ with probability $\min(1,e^{\tau_{rj} + \theta_{rj}[X(r)]_{\cdot j}^T\Omega [X(r)]_{\cdot j}})$\\
			  \textbf{ELSE}\\
			  \indent flip it to $1$ with probability $\min(1,e^{-\tau_{rj}})$.\\
			 Here $\tau_{rj}$ is as defined in \eqref{taupg}. Set$\delta_r^{(t)} =\check{\delta}$.
			
		\end{enumerate}
	\end{enumerate}
\end{algorithm} 
 \section{Simulation studies}
 \label{sim}

 We first present a comparison of the performance of Algorithms \ref{algo:mala} and \ref{algo:polya-gamma} in terms of relative error and time complexity using a  logistic regression with different sample sizes $(n)$ and dimension $(p)$ of the parameter of interest. Secondly we generate data from Ising model with two different structures of $\theta$ and compare the error rates and recovery of the quasi-posterior samples for different data size $(n)$. Lastly to show scalability of the algorithm, we construct credible intervals based on the posterior samples for a network parametrized by a large $300 \times 300$ matrix $\theta$ based on $2000$ observations and check the percentage of active parameters that are covered by the credible intervals. 
  
 \subsection{Comparison of PG and MALA for logistic regression}\label{sec:comp:algo}
\begin{figure}[H]
 	\centering\includegraphics[width = 1\linewidth]{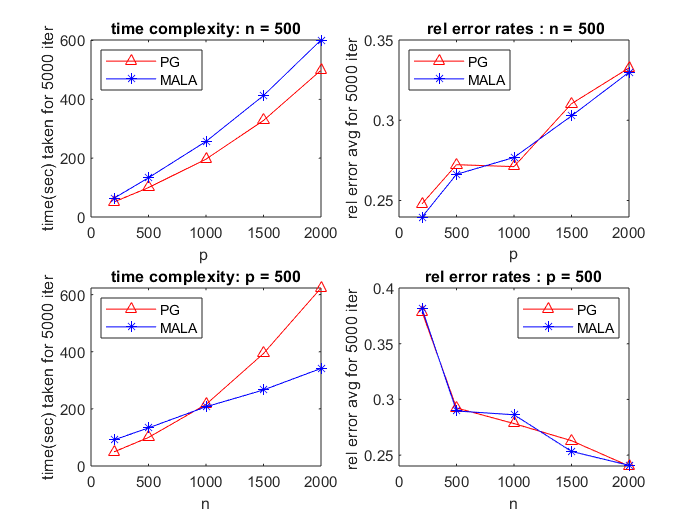}
 	\caption{Comparison of MALA \ref{algo:mala} and PG \ref{algo:polya-gamma} for Logistic Regression based on 5000 iterations}
 	\label{fig:pgmalacom}
 \end{figure}

  We first present results  comparing the two algorithms based on logistic regression in Figure \ref{fig:pgmalacom}. The data was generated based on a parameter $\theta_\star \in \mathbb{R}^p$ which had $10$ active signals of  absolute strength approximately $4\sqrt{10\frac{\log(p)}{n}}$ with a positive or negative sign randomly assigned to them. The regressors were drawn from independent Gaussian distribution and adjusted to have $\|X_j\|_2^2 = n, \ j = 1,\cdots, p$. We used $\rho = \sqrt{\frac{n}{\log(p)}}$, $\gamma = \frac{1}{n\vee p}$ and $u = 2$. We define the relative error and recovery as follows
  \begin{eqnarray}
  	\text{relative error at iteration} \ t: \ e^{(t)} \myeq \frac{||\theta^{(t)} - \theta_{\star}||_2}{\|\theta_{\star}\|_2} \label{eq:error}\\
  	\text{F1 score at iteration} \ t: \ \text{F}1^{(t)} \myeq \frac{2*TA^{(t)}*PA^{(t)}}{TA^{(t)}+PA^{(t)}} \label{eq:rec}
  \end{eqnarray}  
  
  Here, \\
  $TA^{(t)} =$ proportion of true active out of predicted active elements of $\delta$ at iteration $t$ and \\
  $PA^{(t)} =$ proportion of predicted active out of truly active elements of $\delta$ at iteration $t$.\\  
 We run both algorithm for $5000$ iterations. Figure \ref{fig:pgmalacom} shows the relative error (averaged over the number of iterations), as well as the total computation time. We observe that the relative errors from the two samplers remain close. This is expected since the marginal quasi-posterior distribution of $(\delta,\theta)$ is the same in both algorithms. The notable conclusion is that the time complexity for the Polya-Gamma sampler degrades compared to the MaLa sampler when the sample size $n$ is much larger than the dimension. This is due to the fact that sampling  $n$ Polya-Gamma variables at each iteration increases the computation cost of the algorithm significantly. 
 
 \subsection{Numerical experiments using the Ising model}
\begin{figure}[H]
 		\centering
 		\begin{minipage}{.5\textwidth}
 			\centering
 			\includegraphics[width = 1\linewidth]{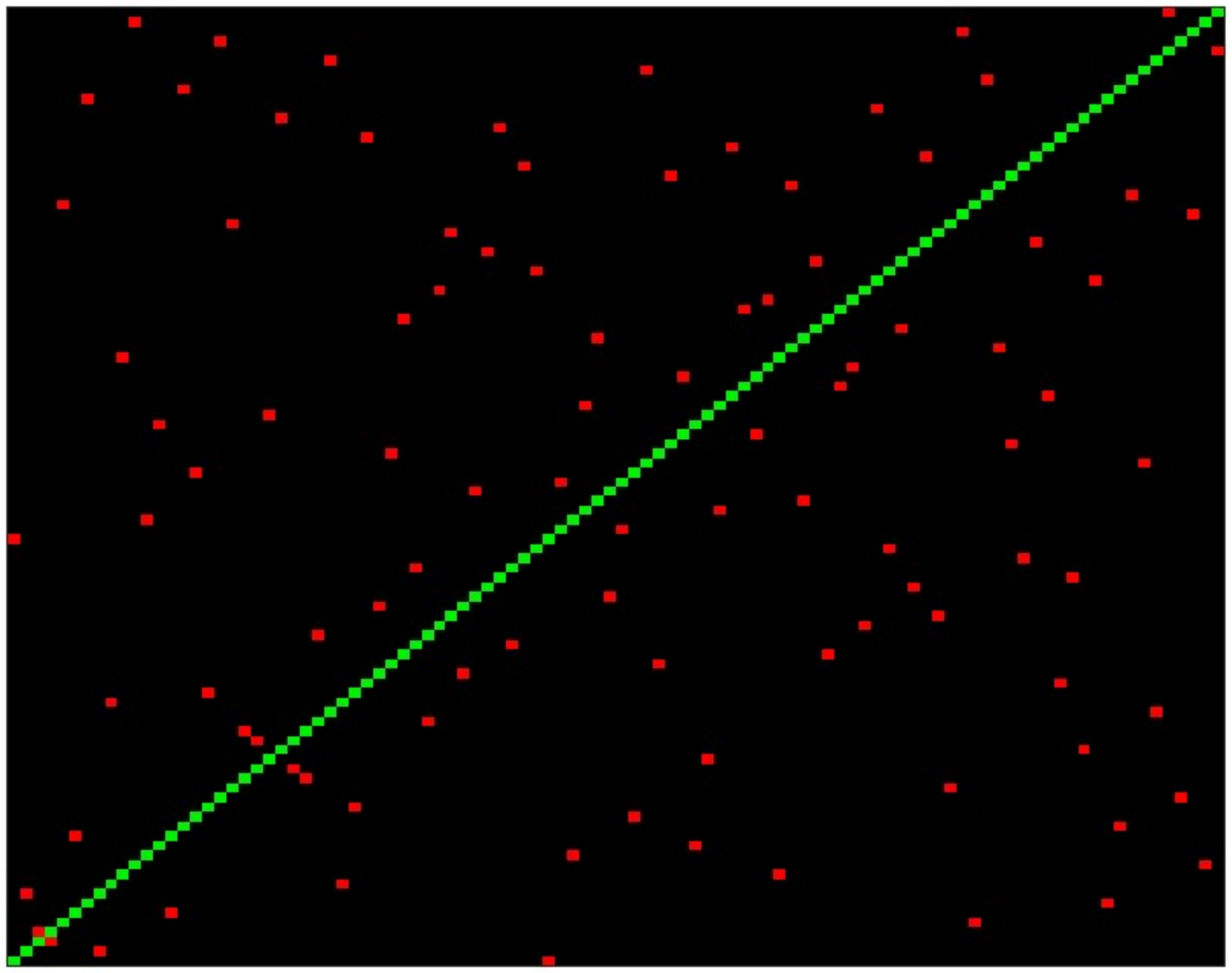}
 			\centering
 			\caption{Heatmap of $\theta_\star$ for network 1}	
 			\label{fig:sim1}
 		\end{minipage}%	
 		\begin{minipage}{.5\textwidth}
 			\centering
 			\includegraphics[width = 1\linewidth]{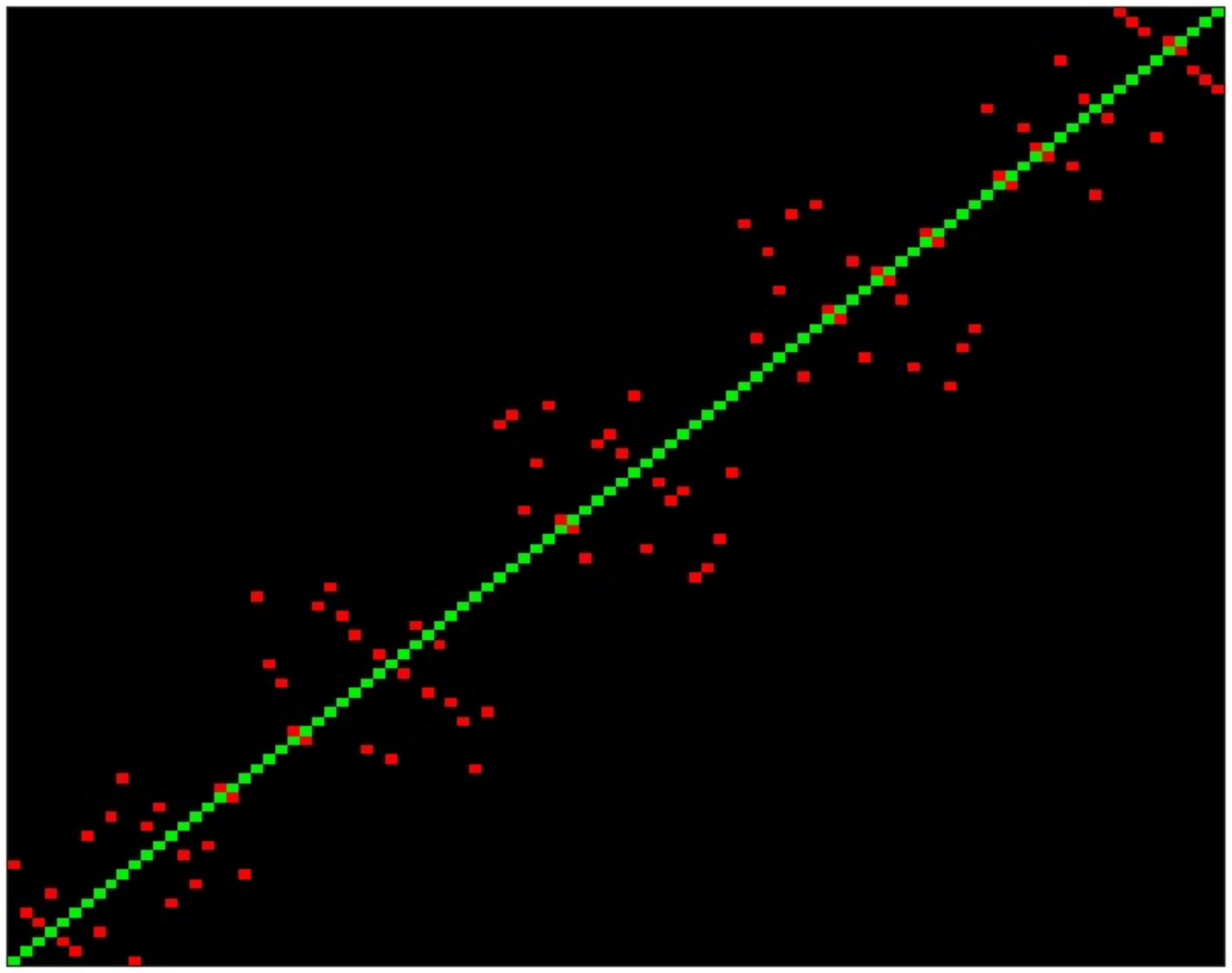}
 			\centering
 			\caption{Heatmap of $\theta_\star$ for network 2}	
 			\label{fig:sim2}
 		\end{minipage}	
 		\small
 		\centering
 		\caption*{The red and green dots indicate positive and negative values of $\theta_{ij}$ respectively}
 \end{figure}

 The next set of results are based on the whole Ising Model. Here we present results based on two networks, one where the structure is completely random (Figure \ref{fig:sim1}) and the other where it consists of clusters along the diagonals (Figure \ref{fig:sim2}). 
 
We introduce the norm $\|\theta\|_0$ as a measure of sparsity where 
\[\|\theta\|_0 = \sum_{r=1}^{p}\sum_{j = 1}^{p} \mathds{1}[\theta_{rj} \ne 0] .\] For each of the two networks, the generating matrix $\theta_\star$ is symmetric in $\mathbb{R}^{100 \times 100}$. Both the networks have 100 non-zero values along the diagonal of $\theta_\star$ and 50 active edges out of 4950 edges, resulting in $\|\theta_\star\|_0 = 200$.

 The Ising model is well known to exhibit  a phase transition phenomenon \cite{georgii}. The phase transition properties of the Ising model may lead to nodes on graph with low or no variability for certain choices of parameter $\theta_\star$ \cite{lizhang:2010}. We carefully chose $\theta_{\star ij}$ to avoid these scenarios. The diagonal elements of  $\theta_\star$ were chosen to be $-2$ and the non-zero off-diagonal $\theta_{\star ij}$'s to be $4$. 
We generate the data from the Ising model using a Gibbs sampler. \\  
The initialization of the MCMC values can be done randomly but the mixing will be much slower in this case. We choose to use the frequentist estimate as initial value at each node $r$ obtained through a proximal gradient descent on the corresponding conditional likelihood \cite{parikh:boyd}. This ensures that the MCMC sampler converges almost immediately. As we noted in Figure \ref{fig:pgmalacom} the PG and MALA sampler produce similar error rates for logistic regression. Hence we present the results of the PG sampler only in case of the Ising Model for the sake of brevity. 

To measure convergence of the MCMC we use the relative error \eqref{eq:error} for each node $r$ referring to then as $e_r^{(t)}$ for the $t_{th}$ iteration and define

\begin{equation*}
	\text{relative error at iteration} \ t \ \text{averaged across nodes}: \ e^{(t)} \myeq \frac{\sum_{r=1}^{p}e^{(t)}_r}{p},
\end{equation*}
Similarly, using \eqref{eq:rec}

\begin{equation*}
	\nonumber\text{F1 score at iteration} \ t  \ \text{averaged across nodes}: \ \text{F}1^{(t)} \myeq \frac{\sum_{r=1}^{p}F1^{(t)}_r}{p}	.	
\end{equation*}

F$1$ score is the combined measure of the power of a method and it's control over false discoveries. A high F1 score indicates low type 1 error and high power. 

\subsection{Behavior of the quasi-posterior distribution with increasing sample size}

 We study here the behavior of the quasi-posterior distribution as the sample size increases. We generate $n$ independent samples from the Ising model with parameter $\theta_\star\in\mathbb{R}^{100\times 100}$, for $n\in\{200,500,1000\}$, where $\theta_\star$ is as described above. Using the simulated data, we ran the PG sampler for 5,000 iterations with $\gamma = \frac{0.1}{p}$, $\rho = \sqrt{\frac{n}{log(p)}}$ and $u = 2$. We initialize the PG sampler using  the frequentist lasso estimate. The relative errors and F1 scores averaged both over the nodes and the last 1,000 iterations are presented in Table \ref{table:tabsize}.  We can see a substantial increase in performance when the sample size grows from 200 to 500 and there is not much gain in terms of precision of estimate as sample size is increased further to 1,000. The quasi-Bayesian approach appears to perform equally well for the two types of network.

\begin{table}[H]
	\centering
	\begin{tabular}{|c|l|c|c|}
		\hline
		& & Average Relative Error & Average F1 score \\
		\hline
		\multirow{3}{*}{\begin{tabular}[c]{@{}l@{}}Network 1 \\ $p = 100$\end{tabular}} &  $n = 200$  & 0.2187 & 0.9336          \\
		& $n = 500$  & 0.0992 & 0.9960                \\
		& $n = 1,000$ & 0.0704& 0.9955  \\
		\hline
		\multirow{3}{*}{\begin{tabular}[c]{@{}l@{}}Network 2\\  $p = 100$\end{tabular}}   & $n = 200$  & 0.1698                 & 0.9689 \\
		& $n = 500$  & 0.0846                 & 1.0000                \\
		& $n = 1,000$ & 0.0690                 & 0.9960    \\         
		\hline
	\end{tabular}
	\caption{Table showing average relative errors and average F1 scores (recovery) for the two networks and different sample sizes.}
	\label{table:tabsize}
\end{table}

\subsection{Behavior of credible intervals for a network with 300 nodes}

\begin{figure}[H]
	\centering
	\begin{minipage}{.7\columnwidth}
		\centering
		\includegraphics[width = \linewidth,height = 7.5cm]{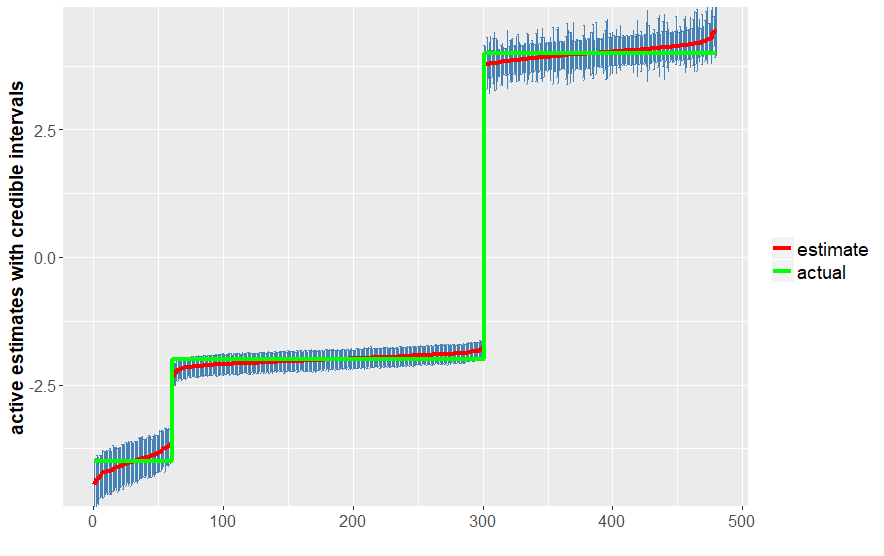}
		\captionsetup{justification=centering,margin=.5cm}
		\caption{Credible intervals of active $\theta_{ij}$ in order of strength of estimates}	
		\label{fig:sim7}
	\end{minipage}	
	\begin{minipage}{.7\columnwidth}
		\centering
		\includegraphics[width = \linewidth]{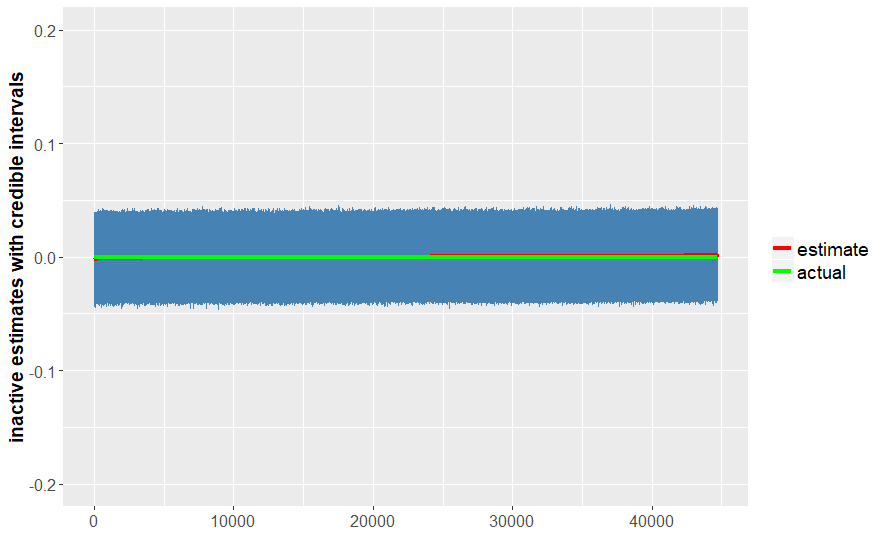}
		\centering
		\captionsetup{justification=centering,margin=.5cm}
		\caption{Credible intervals of inactive $\theta_{ij}$ in order of strength of estimates}	
		\label{fig:sim8}
	\end{minipage}	
	\small
	\centering

\end{figure}

 We generate a larger network  with 300 nodes and 2,000 observations. The network structure is similar to network 2 with block structure along the diagonals but also some sparse active edges along the anti-diagonal resulting in $\max_{r = 1,\cdots,p}\|\delta_{\star r}\|_0 = 3$. Here $\theta_\star$ is symmetric in $\mathbb{R}^{300\times 300}$ with $||\theta_\star||_0 = 660$. The non-zero off-diagonal values of $\theta_\star$ are set at $4$ and the diagonals of $\theta_\star$ are either $-2$ or $-4$. The settings were changed slightly again keeping in mind the phase transition properties of the Ising Model. In this setup, we specifically look at the credible intervals estimated through the MCMC samples using the PG sampler with $\gamma = \frac{0.1}{p}$, $\rho = \sqrt{\frac{n}{\log(p)}}$ and $u = 2$. We run the PG sampler for 30,000 iterations and take the initial 10,000 iterations as burn-in. After the burn-in, the estimates of each $\theta_{ij}$ are obtained by taking the mean of 500 samples, keeping the sample from every $40_{th}$ iteration. The relative error for these 500 samples averaged across the 300 nodes is \textbf{0.0078} while the recovery(F1 score) is calculated to be \textbf{1.0000}. We obtain the final estimate of $\tilde \theta$ after symmetrization of the estimates as mentioned in \eqref{strength}. For the credible interval of $\theta_{\star ij}$ we use the union of the 95\% credible intervals of $\theta_{ij}$ and that of $\theta_{ji}$. Figures \ref{fig:sim7} and \ref{fig:sim8} show the credible intervals of the active and inactive $\theta_{ij}$ separately. We also include the estimates and the true value of the parameter to show the accuracy of the estimates. In 97\% cases the active parameters are covered by the union credible intervals while in 3\% cases they fall just outside. The inactive parameters have credible intervals symmetric around 0. The average credible intervals for each of the 4 distinct true parameter values are given in table \ref{tab:ci} . 
\begin{table}[H]
	\centering
	\begin{tabular}{|c|c|}
		\hline
		True parameter value & Average Credible Interval  \\
		\hline
		0& (-0.037,0.037)\\
		-4& (-4.43,-3.62)\\
		-2&(-2.21,-1.82)\\
		4&(3.66,4.40) \\ 
		\hline
	\end{tabular}
	\caption{Table showing Credible Intervals average for each of the four unique parameter values in the matrix $\theta_\star$ }
	\label{tab:ci}
	\end{table}

The total computing time of our method for this network with 300 nodes and 2000 observations was approximately 600 CPU-hours where each node ran for 30000 iterations. We parallelized the MCMC into 80 parallel processes and the simulation was completed in approximately 8 hours. Given this, we can say that our method is computationally scalable in these data dimensions. 
 
 \newpage
  
 \section{Real data analysis}
 \label{sec:data}
 According to British psychologist Raymond Cattell, variations in human personality is best explained by a model containing sixteen variables (personality factors/traits) \cite{catell:mead}.
 The data that we have analyzed (source: \url{https://openpsychometrics.org/_rawdata/}), comes from an interactive questionnaire of 163 questions designed to measure Cattell's 16 Personality Factors (16PF). For each question, a self-assigned score indicates how accurate it is on a scale of (1) disagree (2) slightly disagree (3) neither agree nor disagree (4) slightly agree (5) agree. Additionally, some other information is collected which includes the test taker's home country, the source from which (s)he got information about the test, her/his perceived accuracy about the answers (s)he provided, age, gender and time elapsed to complete the test. In our analysis, we focused on women in the age group of 30 to 50, who had a self-reported accuracy $\geq$ 75\% and finished the test within half an hour.
 
 The selected data had 4,162 individuals answering 163 questions. Some of the observations had missing values which are represented as 0. The proportion of missing values varied from 0.4\% to 1\% across different questions. The missing values were treated as missing at random and each of them were substituted by a value between 1 to 5. This value was sampled from the marginal distribution of scores for that particular question (covariate).
 
  Table \ref{table:tab1} describes the 16 primary factors. Each factor has 10 questions associated with it except trait B (Reasoning) which has 13 questions leading to a total of 163 questions. 
  
 \begin{table}[H]
 	\centering
 
 \begin{tabular}{ |p{6cm}|p{2cm}| }
% 	\hline
% 	\multicolumn{2}{|c|}{16 PF Primary Factors} \\
 	\hline
 	Trait Name & Trait Code \\
 	\hline
 	Warmth & A \\
 	Reasoning & B\\
 	Emotional Stability & C \\
 	Dominance & E\\
 	Liveliness & F\\
 	Rule-Consciousness & G\\
 	Social Boldness & H\\
 	Sensitivity & I \\
 	Vigilance & L \\
 	Abstractedness & M \\
 	Privateness & N \\
 	Apprehension & O \\
 	Openness to change & Q1 \\
 	Self-reliance & Q2 \\
 	Perfectionism & Q3 \\
 	Tension & Q4 \\
 	\hline
 \end{tabular}
 \caption{16 PF Primary Factors}
 \label{table:tab1}
\end{table}

  We aim to model the network of 163 questions through a Potts model with 163 nodes. Each of the questions are evaluated on a scale of 1 to 5, resulting in a 5-colored Potts model. Our objective is to understand the associations between the questions by estimating the parameter matrix $\theta$ in the Potts model \eqref{PoMo}.We set the coupling function $C(z_r,z_j) = \frac{z_rz_j}{(4)^2}$ and marginal term $C(z_r) = (\frac{z_r}{4})^2$, where $z_r \in (0,1,\cdots, 4)$ after shifting the origin to 0. The denominators in these terms help stabilize the computation of the log-likelihoods and the derivatives required in our MCMC computations. We run the MALA sampler (Algorithm \ref{algo:mala}) using $\rho = \sqrt{n/\log(p)}$, $\gamma = \frac{1}{n}$ and $u = 2$, with a burn-in of $10,000$ iterations. The MCMC runs for $50,000$ more iterations and we keep every $50_{th}$ iteration to obtain a $1,000$ MCMC samples.
 
\begin{figure}[H]
	\centering

		\includegraphics[width = 1\linewidth]{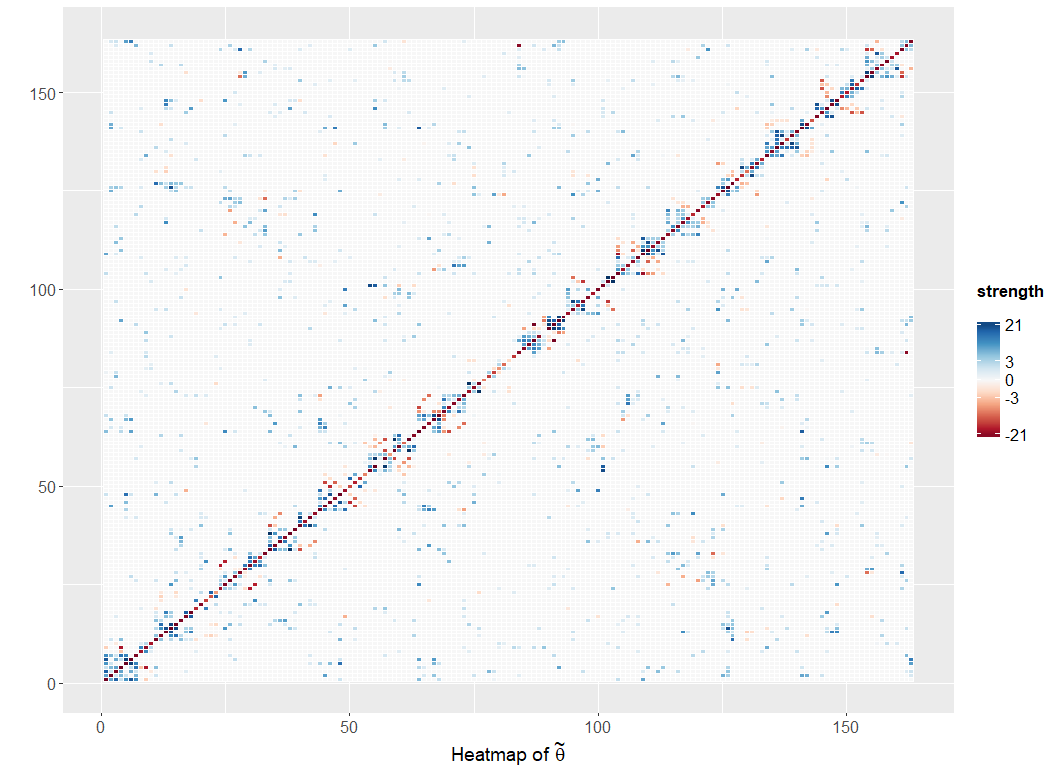}
		\centering
		\caption{The heatmap of $\hat{\theta}$ after symmetrization \eqref{strength}}	
		\label{fig:comp3}

	\small

\end{figure}

	We define 
	\begin{eqnarray}
		\hat{\theta}_{ij} = \frac{1}{1000}\sum_{t =1}^{1000}\theta_{ij}^{(t)}\\
		\hat{P}(\delta_{ij}) = \frac{1}{1000}\sum_{t =1}^{1000}\mathbb{I}(\delta_{ij}^{(t)} = 0)\label{eq:mcmcest}
	\end{eqnarray}
 The final strength of association between node $(i,j)$ based on $1,000$ samples is then measured through a single value $\tilde \theta_{ij}$ evaluated as in \eqref{strength} which has values in the range of $(-21,21)$. The heatmap of the strength of association ($\tilde{\theta}$) is given in Figure \ref{fig:comp3}. The cluster of strong signals around the diagonal represents association between questions relating to the same personality trait while the sparse off-diagonal strong signals represent association between question that are related to two different personality traits. The percentage of estimates with $\hat{P}(\delta_{ij}) = 0$ is around (94\%).

\begin{figure}[H]
	\centering
	\begin{minipage}{.7\columnwidth}
		\centering
		\includegraphics[width = \linewidth,height = 7cm]{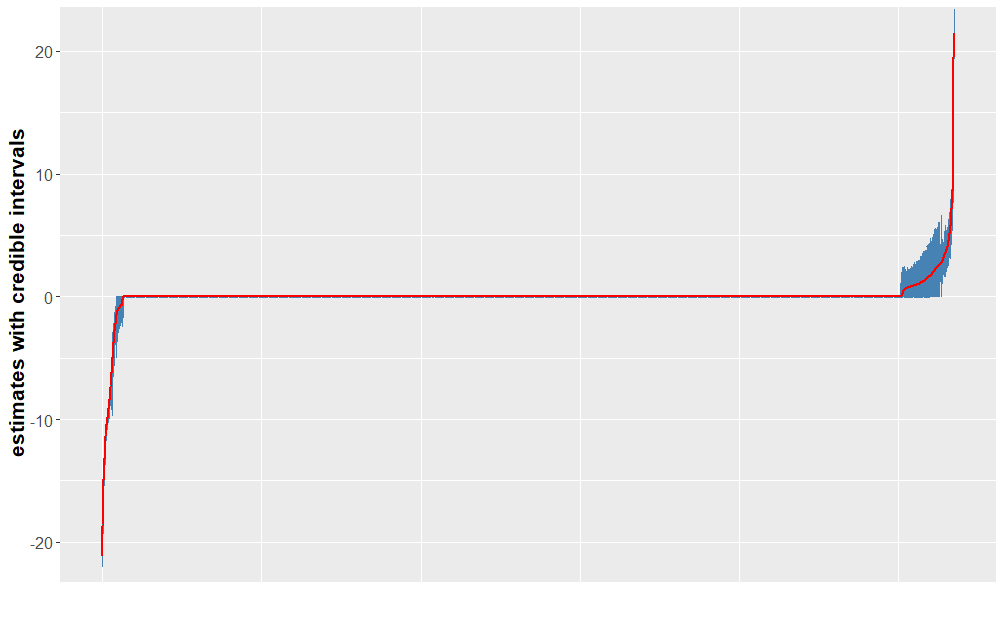}
		\captionsetup{justification=centering,margin=.5cm}
		\caption{ $\tilde{\theta}_{ij}$ (red) with credible intervals (blue) in order of strength of estimates}	
		\label{fig:comp5}
	\end{minipage}	
	\begin{minipage}{.7\columnwidth}
		\centering
		\includegraphics[width = \linewidth]{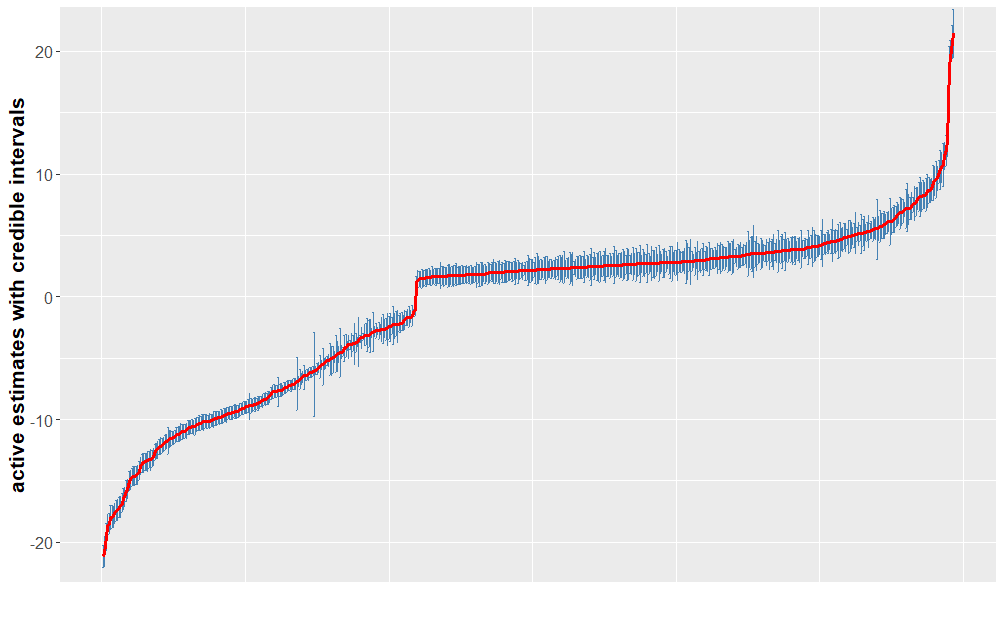}
		\centering
		\captionsetup{justification=centering,margin=.5cm}
		\caption{Fig \ref{fig:comp5} zoomed in for credible intervals not containing 0}	
		\label{fig:comp6}
	\end{minipage}	
	\small
	\centering

\end{figure}

\vspace{4mm}
The credible region for the estimate of $\theta_{ij}$ are evaluated as union of the 95\% credible intervals of $\theta_{ij}$ and $\theta_{ji}$, obtained from the respective set of MCMC samples. Figure \ref{fig:comp5} shows the estimated credible intervals for all the parameters ($\theta_{ij}$). It demonstrates the fact that for most inactive parameters the credible set is a very small interval around 0 which given the scale of the image appears as a straight line. Figure \ref{fig:comp6} is a zoomed in version of Figure \ref{fig:comp5} corresponding to parameters whose credible intervals do not contain 0. 

\begin{figure}[H]
  	\centering
  	\includegraphics[scale = .7]{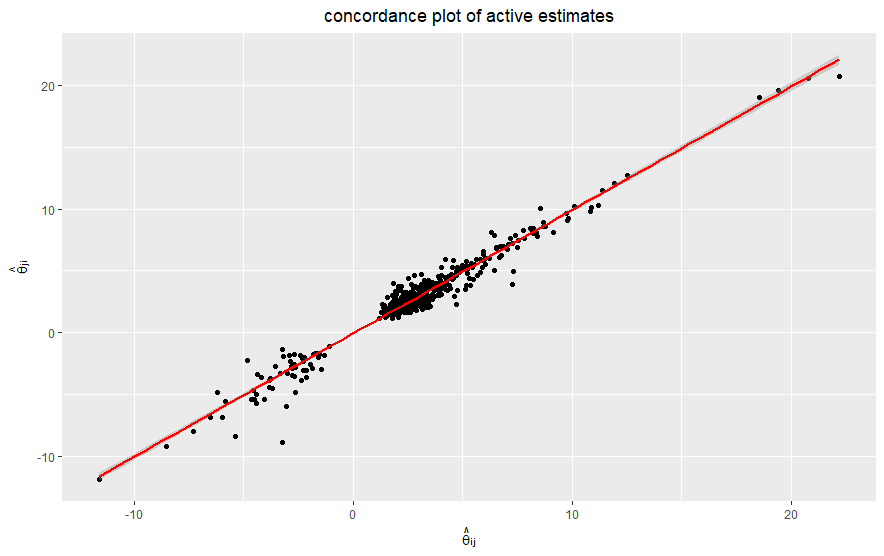}
  	\caption{concordance plot for estimates with credible intervals not containing 0. Fitted line in red has intercept: -0.08 and slope: 0.9995 }
  	\label{fig:comp7}
  \end{figure}

We introduce Figure \ref{fig:comp7} to show the concordance between the estimates $\hat{\theta}_{ij}$ and  $\hat{\theta}_{ji}$ for those estimates whose union credible intervals do not contain 0. The figure shows a high level of concordance.

 \cite{catell:mead} used several techniques including factor analysis to establish that personality structure is hierarchical, with primary and secondary level traits. The primary level consists of the 16 personality traits (used in our analysis). The secondary level consists of a version of the Big Five Traits corresponding to broader human qualities. They are obtained by factor-analyzing the correlation matrix of the 16 primary-level personality traits. 
  
The grouping of the 16 primary factors into the Big Five Traits are shown in Table \ref{table:tab2}. Reasoning (trait B) stands alone without any association to the Big Five Traits.
   \begin{table}[H]
   		\centering
   		
   	\begin{tabular}{ |p{6cm}|p{6cm}| }
   	   		% 		\hline
   		% 		\multicolumn{2}{|c|}{Association of the 16 traits with the five broader human characteristics} \\
   		\hline
   		Big Five Traits & Associated 16PF Traits \\
   		\hline
   		Introversion/Extroversion & A, F, H, N, Q2 \\
   		Low anxiety/High Anxiety & C, L, O, Q4\\
   		Receptivity/Tough-Mindedness & A, I, M, Q1 \\
   		Accommodation/Independence & E, H, L , Q1\\
   		Lack of Restraint/Self Control & F, G, M, Q3\\
   		-- & B \\
   		\hline
   	\end{tabular}
   	\caption{Grouping of the 16 primary factors into the Big Five Traits}
   	\label{table:tab2}
   \end{table}
    
 With the results of the analysis we now wish to see if the 16 primary factors show similar associations as the ones established in Table \ref{table:tab2}, thus providing a validation to the inference. In order to do so, we start  the probability of edge between the questions $(i,j)$ given by $\tilde p_{ij}$ \eqref{edge} and \eqref{eq:mcmcest}. We summarize the estimates of probability of edge between 163 questions into a smaller $16\times 16$ matrix $\phi$ corresponding to the 16 traits. We define the set $S_i = \{\text{questions under trait} \ i\}$ and $n_{ij}$ to be the total number of possible edges between trait $i$ and trait $j$. We define the matrix $\phi$ as
 \begin{align*}
 \phi_{ij}  = & \frac{1}{n_{ij}}\sum_{k \in S_i, l \in S_j} \tilde p_{kl} \quad .
 \end{align*}

 The off-diagonal elements of the matrix $\phi$ measure the average probability of association between each pair of traits. The element-wise reciprocal of this matrix gives us a pseudo-distance measure between the 16 traits which is used to form a hierarchical clustering using Ward's method (ward.D2 in \emph{stats:hclust} in R). 

\begin{figure}
  	\centering
  	\includegraphics[scale = .75]{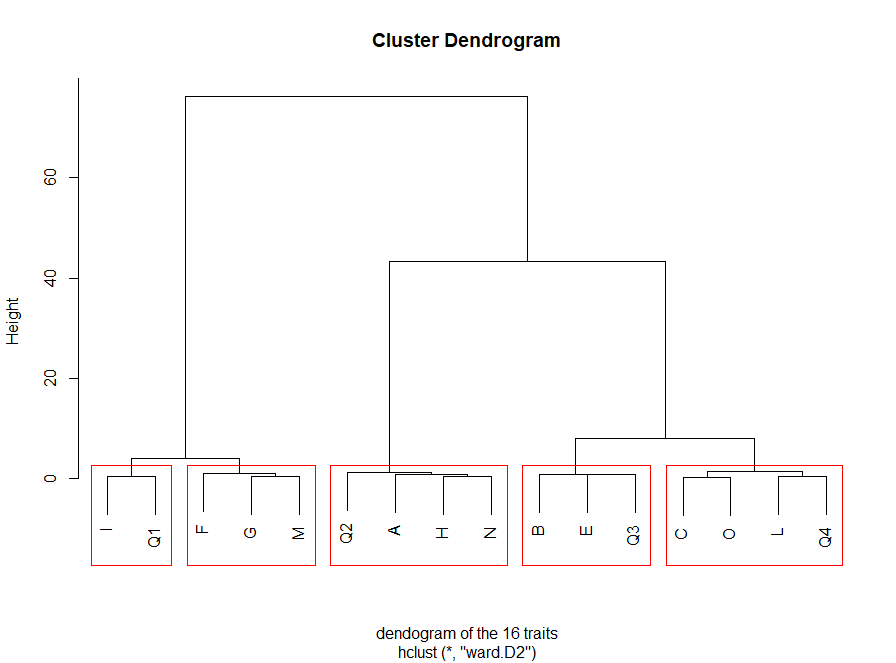}
  	\caption {dendogram identifying clusters of the 16 traits}
  	\label{fig:comp8}

  \end{figure}

 Figure \ref{fig:comp8} shows the results of the clustering. We see that our method perfectly recovers the low-anxiety/high-anxiety (C,L,O,Q4) cluster [Table \ref{table:tab2} ]. It also nearly recovers Introversion/Extroversion (A, H, N, Q2)[Table \ref{table:tab2} ]. The trait F(liveliness) [Table \ref{table:tab1} ] which is common to both Introversion/Extroversion and Lack of Restraint/Self-Control in Table \ref{table:tab2} is shown to be clustered more strongly with the later group and we also recover most of the Lack of Restraint/Self Control Cluster (F,G,M). In our clustering (I,Q1) are also placed together which is substantiated by the fact that they are common to the Receptivity/Tough-Mindedness cluster [Table\ref{table:tab2} ]. Additionally we find that given the data and the demographics with which we chose to work our method identifies a new cluster (E,Q3,B) which may lead to possible novel insights for this particular demographic warranting further investigations. Thus we see that several groupings in Table \ref{table:tab2} corresponding to the Big Five Traits are reflected in our method.

  \section{Conclusion}
  \label{sec:conc}
  
  In this article, we have developed a quasi-Bayesian approach to fit  large  Potts models. The use of a pseudo-likelihood  and a prior distribution that factorizes across  the columns of the parameter matrix has enabled us to side-step the intractable normalization constant of the Potts model and perform computations in parallel for each node of the graph. 
  We have shown in our simulations that for appropriate choices of the hyper-parameters, the  method recovers the true data-generating parameters and  achieves high F1 scores for moderate sample size. We have also shown that the proposed MCMC algorithms can easily handle problem sizes up  to $p = 300$, and possibly more if access to a computer with a large number of cores is available.
  Finally we have implemented the method on a 16 Personality Factor dataset and shown through a hierarchical clustering of our estimates that some of the important features of association between the 16 Personality Factors in captured in our estimates thus validating our inference. The proposed quasi-Bayesian approach has also good theoretical properties as explored in \cite{atchade:2019}.

\noindent

 \bigskip
 \begin{center}
 	
 	{\large\bf SUPPLEMENTARY MATERIAL}
 \end{center}
 
 \begin{description}
 	 	
 	\item[Matlab-code:] The matlab package with readme file is located at \url{https://github.com/anweshaumich/DGM_parcomp} 
 	
 	\item[16PF data set:] The Data set used for analysis in section \ref{sec:data} can be obtained from \url{https://github.com/anweshaumich/DGM_parcomp/blob/master/16PF_data_used.zip}. The data has been filtered to consider women between age between 30 to 50, self-reported accuracy of 75 or above and completion time less than half hour ("smalldat.csv"). After adjusting for the missing data we have the dataset "distmiss.csv".

 \end{description}
\singlespacing 	
\bibliographystyle{Chicago} 	
\bibliography{ref,ref2}
\end{document}